\documentclass[12pt]{article}

\usepackage{mathrsfs}
\usepackage[T1]{fontenc}
\usepackage{mathpazo}
\usepackage{setspace}
\usepackage{amsfonts}
\usepackage{amssymb}
\usepackage{amsmath}
\usepackage{esint}                   
\usepackage{epsfig}
\usepackage{latexsym}
\usepackage{color}
\usepackage{graphicx}
\usepackage{hyperref}
\usepackage{nicefrac}
\usepackage[latin1]{inputenc}
\usepackage{pstricks}
\usepackage{slashed}
\usepackage{multirow}
\usepackage{ulem}
\usepackage{comment}
\usepackage[nosort]{cite}
\usepackage{datetime}
\usepackage{tikz-cd}
\usepackage{float}

\def\hybrid{\topmargin -30pt    \oddsidemargin 0pt 
        \headheight 0pt \headsep 0pt
        \textwidth 6.25in       
        \textheight 9.5in       
        \marginparwidth .875in
        \parskip 5pt plus 1pt   \jot = 1.5ex}

\hybrid

\def\baselinestretch{1.2}

\catcode`\@=11

\def\marginnote#1{}
\newcount\hour
\newcount\minute
\newtoks\amorpm
\hour=\time\divide\hour by60
\minute=\time{\multiply\hour by60 \global\advance\minute by-\hour}
\edef\standardtime{{\ifnum\hour<12 \global\amorpm={am}%
        \else\global\amorpm={pm}\advance\hour by-12 \fi
        \ifnum\hour=0 \hour=12 \fi
        \number\hour:\ifnum\minute<10 0\fi\number\minute\the\amorpm}}
\edef\militarytime{\number\hour:\ifnum\minute<10 0\fi\number\minute}

\def\draftlabel#1{{\@bsphack\if@filesw {\let\thepage\relax
   \xdef\@gtempa{\write\@auxout{\string
      \newlabel{#1}{{\@currentlabel}{\thepage}}}}}\@gtempa
   \if@nobreak \ifvmode\nobreak\fi\fi\fi\@esphack}
        \gdef\@eqnlabel{#1}}
\def\@eqnlabel{}
\def\@vacuum{}
\def\draftmarginnote#1{\marginpar{\raggedright\scriptsize\tt#1}}

\def\draft{\oddsidemargin -.5truein
        \def\@oddfoot{\sl preliminary draft \hfil
        \rm\thepage\hfil\sl\today\quad\militarytime}
        \let\@evenfoot\@oddfoot \overfullrule 3pt
        \let\label=\draftlabel
        \let\marginnote=\draftmarginnote
   \def\@eqnnum{(\theequation)\rlap{\kern\marginparsep\tt\@eqnlabel}%
\global\let\@eqnlabel\@vacuum}  }

\def\draft2{
        \def\@oddfoot{\sl preliminary draft \hfil
        \rm\thepage\hfil\sl\today\quad\militarytime}
        \let\@evenfoot\@oddfoot \overfullrule 3pt
        \let\marginnote=\draftmarginnote
   \def\@eqnnum{(\theequation)\rlap{\kern\marginparsep\tt\@eqnlabel}%
\global\let\@eqnlabel\@vacuum}  }

\def\preprint{\twocolumn\sloppy\flushbottom\parindent 2em
        \leftmargini 2em\leftmarginv .5em\leftmarginvi .5em
        \oddsidemargin -.5in    \evensidemargin -.5in
        \columnsep .4in \footheight 0pt
        \textwidth 10.in        \topmargin  -.4in
        \headheight 12pt \topskip .4in
        \textheight 6.9in \footskip 0pt
        \def\@oddhead{\thepage\hfil\addtocounter{page}{1}\thepage}
        \let\@evenhead\@oddhead \def\@oddfoot{} \def\@evenfoot{} }

\def\numberbysection{\@addtoreset{equation}{section}
        \def\theequation{\thesection.\arabic{equation}}}

\def\underline#1{\relax\ifmmode\@@underline#1\else
        $\@@underline{\hbox{#1}}$\relax\fi}

\def\titlepage{\@restonecolfalse\if@twocolumn\@restonecoltrue\onecolumn
     \else \newpage \fi \thispagestyle{empty}\c@page\z@
        \def\thefootnote{\fnsymbol{footnote}} }

\def\endtitlepage{\if@restonecol\twocolumn \else \newpage \fi
        \def\thefootnote{\arabic{footnote}}
        \setcounter{footnote}{0}}  

\catcode`@=12
\relax

\def\figcap{\section*{Figure Captions\markboth
        {FIGURECAPTIONS}{FIGURECAPTIONS}}\list
        {Figure \arabic{enumi}:\hfill}{\settowidth\labelwidth{Figure
999:}
        \leftmargin\labelwidth
        \advance\leftmargin\labelsep\usecounter{enumi}}}
 \relax
\def\tablecap{\section*{Table Captions\markboth
        {TABLECAPTIONS}{TABLECAPTIONS}}\list
        {Table \arabic{enumi}:\hfill}{\settowidth\labelwidth{Table
999:}
        \leftmargin\labelwidth
        \advance\leftmargin\labelsep\usecounter{enumi}}}
 \relax
\def\reflist{\section*{References\markboth
        {REFLIST}{REFLIST}}\list
        {[\arabic{enumi}]\hfill}{\settowidth\labelwidth{[999]}
        \leftmargin\labelwidth
        \advance\leftmargin\labelsep\usecounter{enumi}}}
 \relax
\makeatletter
\newcounter{pubctr}
\def\publist{\@ifnextchar[{\@publist}{\@@publist}}
\def\@publist[#1]{\list
        {[\arabic{pubctr}]\hfill}{\settowidth\labelwidth{[999]}
        \leftmargin\labelwidth
        \advance\leftmargin\labelsep
        \@nmbrlisttrue\def\@listctr{pubctr}
        \setcounter{pubctr}{#1}\addtocounter{pubctr}{-1}}}
\def\@@publist{\list
        {[\arabic{pubctr}]\hfill}{\settowidth\labelwidth{[999]}
        \leftmargin\labelwidth
        \advance\leftmargin\labelsep
        \@nmbrlisttrue\def\@listctr{pubctr}}}
 \relax
\makeatother

\def\be{\begin{equation}}
\def\ee{\end{equation}}
\def\ba{\begin{eqnarray}}
\def\ea{\end{eqnarray}}

\def\del{\partial}

\def\k{\kappa}
\def\r{\rho}
\def\a{\alpha}

\def\b{\beta}

\def\g{\gamma}
\def\G{\Gamma}
\def\d{\delta}
\def\D{\Delta}
\def\e{\epsilon}

\def\p{\pi}

\def\th{\theta}

\def\m{\mu}
\def\n{\nu}

\def\l{\lambda}

\def\s{\sigma}

\def\cA{{\cal A}}

\def\cH{{\cal H}}
\def\cL{{\cal L}}

\def\no{\noindent}

\def\qq{\qquad}

\def\IR{\relax{\rm I\kern-.18em R}}

\def\inv{^{\raise.0ex\hbox{${\scriptscriptstyle -}$}\kern-.05em 1}}

\def \ha {{\frac{1}{2}}}

\def \ov {\over}

\def\diag{{\rm diag}}

\begin{document}


\renewcommand{\theequation}{\thesection.\arabic{equation}}
\csname @addtoreset\endcsname{equation}{section}

\begin{titlepage}
\begin{center}

\hfill CERN-TH-2022-129

\renewcommand*{\thefootnote}{\arabic{footnote}}

\phantom{xx}
\vskip 0.5in

{\large {\bf Scattering in integrable pp-wave backgrounds: \\
\vskip .07 cm
$S$-matrix and absence of particle production }}

\vskip 0.5in

{\bf George Georgiou${}^1$}\hskip .1cm and
{\bf Konstantinos Sfetsos}${}^{1,2}$

\vskip 0.11in

${}^1$ Department of Nuclear and Particle Physics, \\
Faculty of Physics, National and Kapodistrian University of Athens, \\
Athens 15784, Greece

\vskip 0.1in

${}^2$ Theoretical Physics Department,
 CERN, 1211 Geneva 23, Switzerland

\vskip .3 cm

{\footnotesize \texttt george.georgiou, ksfetsos@phys.uoa.gr}


\vskip .2in

\end{center}

\vskip .2in

\centerline{\bf Abstract}
\no
Particle production in integrable field theories may exist depending on the vacuum around which 
excitations are defined. 
To tackle this and analogous issues with conventional field theoretical tools, we consider the integrable $\l$-deformed model for $SU(2)$ together with a timelike coordinate. We construct the corresponding four-dimensional plane wave background keeping also post-plane wave corrections, as well as all the non-trivial $\l$-dependence. 
 After imposing  the light-cone gauge and the Virasoro constraints, we obtain an interacting field theory for the transverse physical modes which are massive.
 We explicitly demonstrate the absence of particle production to leading order in 
 the large $k$-expansion. This is based crucially on the form of the interaction vertices and their dependence 
 on the $\l$-deformation parameter.  In addition, we compute the $S$-matrix for the two-particle elastic scattering exactly in $\l$ and to leading order in the large $k$-expansion. Our method can be applied to any integrable theory with at least one isometry.

\vskip .4in

\vfill

\end{titlepage}
\vfill
\eject



\def\baselinestretch{1.2}
\baselineskip 20 pt

\newcommand{\eqn}[1]{(\ref{#1})}

\tableofcontents


\section{Introduction}

Non-linear $\s$-models in 1+1 spacetime dimensions are of  particular interest because they often represent the worldsheet  of string theories. Among $\s$-models, the class of integrable ones is of great importance. At the classical level, integrability is established by rewriting the classical equations of motion as the flatness condition of a Lax pair that depends on a spectral parameter. This rewriting guarantees the existence of infinitely many conserved charges which in turn ensures classical integrability, when  these charges are in involution. 

However, not all classically integrable models remain integrable at the quantum level. In order to prove quantum integrability  one should demonstrate that scattering of particles is non-diffractive. In particular, this means that no particles are being created or annihilated in a scattering process and that the complete S-matrix of the theory is fully determined by the two-body S-matrix. Two important examples which are believed to retain integrability at the quantum level are the principal chiral model (PCM) \cite{Luscher:1977uq,Goldschmidt:1980wq,Ogievetsky:1987vv} and certain deformations of it that go under the name of $\l$-models \cite{Sfetsos:2013wia, Hollowood:2014rla}. 
The simplest 
class of these  $\l$-deformed models is based on a group $G$ and interpolates between a WZW model at level $k$ and the non-abelian T-dual of the PCM \cite{Sfetsos:2013wia} (the $SU(2)$ case was constructed before with different 
methods in \cite{Balog:1993es}). More general integrable  $\l$-models based on products of semi-simple groups were constructed in \cite{Georgiou:2016urf,Georgiou:2017jfi,Georgiou:2018hpd,Georgiou:2018gpe}, while the most generic models of this type were constructed in \cite{Georgiou:2020wwg,Georgiou:2021pbd}. These models exhibit a very interesting and rich renormalisation group (RG) behaviour, namely they flow from a sum of WZW models in the ultraviolet (UV) to certain 2-dimensional conformal field theories (CFTs) in the infrared (IR).\footnote{Their quantum structure, including correlation functions of currents and primary operators as exact functions of the deformation parameters, was studied in
\cite{Itsios:2014lca, Sfetsos:2014jfa,Georgiou:2015nka,Georgiou:2016iom,Georgiou:2016zyo,Georgiou:2017oly,Georgiou:2019jcf,Georgiou:2019nbz}.}

In the case of the PCM, integrability persists at the quantum level and the two-particle S-matrix acquires the characteristic product form
\be\label{S-PCM}
S(\th)=f(\th)S_{G_L}(\th)\otimes S_{G_R}(\th),\qquad \th= \th_1-\th_2\, ,
\ee
where $S_{G(\th)}$ is a $G$-invariant building block of the complete S-matrix. It is built from a rational solution of the Yang-Baxter (Y-B) equation. Furthermore, $f(\th)$ is a scalar function accounting for the poles of the S-matrix related to the bound states, while $\th_1$ and $\th_2$ denote the rapidities of the incoming particles. This factorised form of the S-matrix reflects the $G_L \otimes G_R$ symmetry of the model.\\
In the case of the simplest $\l$-model with group $SU(2)$, the deformation breaks the $SU(2)_R$ symmetry. Rather than being completely lost, $SU(2)_R$, or more precisely its Yangian extension,  is deformed to an affine quantum group ${\cal U}_q(SU(2))$ with the deformation parameter being $q=e^{i \pi/(k+2)}$, where $k$ is the WZW level.
The exact S-matrix of this model,
proposed in \cite{Ahn:1990gn,Evans:1994hi,Hollowood:2015dpa,Appadu:2017fff,Appadu:2017bnv}, takes again the product form \eqref{S-PCM} but with the right part of the S-matrix replaced by the restricted solid on solid (RSOS) S-matrix which appears in the context of the restricted sine-Gordon theory \cite{LeClair:1989wy,Bernard:1990ys,Bernard:1990cw} and which is invariant under the aforementioned affine quantum group, namely
\be\label{lambda}
S_\l(\th)=S_{SU(2)_L}(\th)\otimes S_{RSOS}(\th)\, .
\ee
Two important comments are in order. Firstly, the presence of the RSOS part of the S-matrix implies that the states which are scattered are solitonic states carrying kink quantum numbers. Secondly, the conjectured S-matrix of \eqref{lambda} is an exact  S-matrix that depends only on  $k$ and the difference of the rapidities but does not depend at all on the deformation parameter $\l$.

The above quantum S-matrix proposals, despite being exact, have the disadvantage that they can not be directly connected, let alone derived, from the Lagrangian of the model. This is so because the objects which scatter are massive objects with their mass $m$ being dynamically generated in the IR regime. 
Only indirect consistency checks of these proposals can be made. Based on the work of \cite{Polyakov:1983tt,Wiegmann:1984pw}, various authors performed  consistency checks of these S-matrix proposals for several groups\cite{Hasenfratz:1990zz,Balog:1992cm,Fateev:1994ai,Hollowood:1994np}. One may introduce an external field $h$ and calculate the response of the free energy of the system when $h\gg m$ by employing TBA techniques. Then one can use one loop perturbation theory to evaluate the free energy when $h\gg \Lambda$, where  $\Lambda$ is the quantity analogous to the QCD scale $\Lambda_{QCD}$. The two expressions for the free energy have similar functional forms and they completely agree if one demands that the ratio ${m \ov \Lambda}$ to be a specific function of $k$. Consequently, this calculation determines the mass gap of the theory. For the $\lambda$-model with $SU(2)$ group this calculation was performed in \cite{Evans:1994hi}. 
We note that, initial calculations of the mass gap using perturbative methods were performed in \cite{Luscher:1982uv, Floratos:1984bz}.

One may try to circumvent the aforementioned disadvantage by expanding the action of a generic integrable theory around the trivial vacuum, which, for $\s$-models based on group manifolds, is at $g={\mathbb 1}$. For the single $\l$-models this was done in \cite{Georgiou:2019aon}, with the resulting action  being very effective for addressing the calculations of $\beta$-functions and anomalous dimensions of operators. However, the S-matrix obtained from this action does not satisfy the conditions of the absence particle production and factorization \cite{Hoare:2018jim}. 
Indeed, the connection to integrability is lost,  already at the tree-level \cite{Nappi:1979ig,Figueirido:1988ct}, in the case where the spectrum contains massless excitations.  The fact that the massless S-matrix of classically integrable theories exhibits particle production is known for a long time but not quite appreciated. The standard argument that integrability implies factorization and the absence of particle production is known to hold only for the non-perturbative massive S-matrix \cite{zam-zam} and formally applies only to the massive case \cite{Parke:1980ki,Shankar}.

In this work, we develop a method which re-establishes the link between integrability and factorisation of the S-matrix. This is accomplished by expanding around a vacuum which supports massive excitations. Our method can be applied to any integrable theory with one or more isometries. One adds to the $\sigma$-model action a timelike
spectator field $t$  which plays the role of time. One then takes the Penrose limit of the geometry around a null geodesic involving the isometry of the initial background and $t$. Subsequently, one fixes the light-cone gauge and imposes the Virasoro constraints so that he is left only with the transverse degrees of freedom. All transverse excitations are massive and the properties of factorisation and no particle production can now be unambiguously addressed. A similar method has already been applied in the context of superstring theories \cite{Klose}.

We exemplify our method by considering the simplest possible example, that of the isotropic $\l$-model with group $SU(2)$. In section  \ref{pp-post}, we calculate the pp-wave background of the aforementioned model, as well as the post pp-wave corrections which will account for the interactions of the massive excitations. In section \ref{Lagr}, we choose the light-cone condition and impose the Virasoro constraints. In this way, one is left with two massive propagating degrees of freedom. We present the Lagrangian governing the propagation of the physical modes up to ${\cal O}(1/k^{3/2})$. It inherits a certain non-perturbative symmetry which is present in the original $\l$-model. In section \ref{secS-matrix}, we derive the S-matrix of our model and show that it does not exhibit particle production up to $ {\cal O}(1/k^{3/2})$. Furthermore, one can easily show that it satisfies the Yang-Baxter equation up to the same order. Finally, in section \ref{conclusions} we draw our conclusions. We close the paper with two appendices. In appendix  \ref{AppA}, we consider various pp-wave limits related to our model which may prove useful in further investigations, while in appendix \ref{AppB}, we prove that the light-cone gauge fixing is consistent with the equation of motion of $x^-$, as long as the curved worldsheet metric is properly chosen.

\section{The plane wave and post-plane wave corrections}
\label{pp-post}

Two-dimensional  $\s$-models are described in terms of the 
action
\begin{equation}
\label{action}
S= \frac{1}{2\pi}\int d^2\sigma \cL\ ,\qq
 \cL = (G_{\m\n}+ B_{\m\n}) \del_+x^\m\del_-x^\n\,,
\end{equation}
from which one can read off the metric and the  antisymmetric tensor fields. 
Our conventions for the world-sheet coordinates $\s^\pm$ and $(\tau,\s)$ are given below
\be
\s^\pm=\tau\pm\s\,,\quad \del_\pm=\frac12\left(\del_\tau\pm\del_\s\right)\,,\quad \text{d}^2\s=d\tau\,d\s\,.
\ee

\no
With the above normalization of fields we specialise to the $\l$-deformation for the $SU(2)$ case. 
We have the metric \cite{Sfetsos:2013wia}
\be
\label{g.su2}
\begin{split}
&ds^2=2k\left(\frac{1+\l}{1-\l} d\a^2+\frac{1-\l^2}{\D(\a)}\sin^2\a\left(d\beta^2+\sin^2\beta d\gamma^2\right)\right)\,,\\
&\D(\a)=(1-\l)^2\cos^2\a+(1+\l)^2\sin^2\a\,
\end{split}
\ee
and the antisymmetric tensor
\be
\label{B.su2}
B=2k\left(-\a+\frac{(1-\l)^2}{\D(\a)}\sin\a\cos\a\right)\sin\beta\, d\beta\wedge d\gamma\, .
\ee
The physical range of the parameter $\l$ is $0\leqslant \l <1$ follows from the construction
 \cite{Sfetsos:2013wia}. 
The above background is invariant under the non-perturbative symmetry
\be
\label{syymm}
\l\to {1\ov \l}\ ,\qq k\to -k \ ,\qq \a\to - \a\ ,
\ee
where the flip in the sign of $\a$ induces an inversion in the $SU(2)$ group element.
The symmetry \eqref{syymm} is in accordance with the symmetry 
of $\l$-deformed $\s$-models based on  a general group for which the symmetry is implemented be sending $g\rightarrow g^{-1}$
 \cite{Itsios:2014lca}. For $\l=0$ we above background fields correspond to the WZW model for the group $SU(2)$ at level $k$.

To the above metric \eqref{g.su2}, we add a term containing a time coordinate $t$. Conveniently, we parametrise it  as $\displaystyle - 2 k {1 - \l\ov 1 + \l} dt^2$. 
The resulting four-dimensional spacetime has then an obvious null geodesic given by
\be 
\a=\b={\pi\ov 2}\ ,\qq t=\g\ .
\ee
Consider, now, the following change of variables
\be
\begin{split}
& t = \ha \sqrt{1+\l\ov 1-\l} \bigg(x^+ - {x^-\ov k}\bigg)\ ,\qq \g = \ha \sqrt{1+\l\ov 1-\l} \bigg(x^+ + {x^-\ov k}\bigg)\ ,
\\
&
\a = {\pi\ov 2} + \sqrt{1-\l\ov 2 k(1+\l)}\, x_1\ ,\qq  \b = {\pi\ov 2} + \sqrt{1+\l\ov 2 k(1-\l)}\, x_2\ .
\end{split}
\ee
In the limit $k\gg 1$, the Lagrangian density should have  an expansion of the form 
\be
\label{lagrexp}
\cL = \cL^{(0)}  + {1\ov k}\, \cL^{(1)}  + {\cal O}(1/k^2)  \ ,
\ee
where we note that in our case additional terms with fractional powers of $k$, e.g. of ${\cal O}(1/k^{1/2})$ and  ${\cal O}(1/k^{3/2})$, contribute only to the antisymmetric tensor with vanishing however field strength. 
Therefore we have safely omitted them. As a result, all our calculations are valid up to, including, order ${\cal O}(1/k^{3/2})$.

The first term in \eqref{lagrexp} has a metric and an antisymmetric tensor given by
\be
\label{ppw}
\begin{split}
& ds^{(0)2}  = 2 d  x^+ d x^- + dx_1^2 +  dx_2^2
 -{1\ov 4} \bigg( \Big({1-\l\ov 1+\l}\Big)^3 x_1^2 + {1+\l\ov 1-\l} x_2^2\bigg) (dx^+)^2 \ ,
\\
&
B^{(0)} = -{1+\l^2\ov \sqrt{(1-\l)(1+\l)^3}}\, x_1\, dx_2\wedge dx^+
\end{split}
\ee
and represents a plane wave expressed in its Brinkman form.
 
\no
The interactions are encoded in the ${\cal O}(1/k)$ terms with the corresponding metric and antisymmetric tensor corrections given by
\be
\label{ppww}
\begin{split}
& ds^{(1)2} =  
  -{1\ov 2} \bigg( \Big({1-\l\ov 1+\l}\Big)^3 x_1^2 + {1+\l\ov 1-\l} x_2^2\bigg) dx^+ dx^-
-\ha \Big({1-\l\ov 1+\l}\Big)^3 x_1^2\, dx_2^2
\\
& \qq + {1\ov 24} \Bigg[ {(1-\l)^4 (1-10 \l + \l^2)\ov (1+\l)^6} x_1^4 + \Big({1+\l\ov 1-\l}\Big)^2x_2^4 + 3 \Big({1-\l\ov 1+\l}\Big)^2 x_1^2 x_2^2 \Bigg] (dx^+)^2
\\
& B^{(1)}= \bigg({1\ov 4} {1+\l^2\ov \sqrt{(1-\l)^3(1+\l)}}\, x_1 x_2^2 + {1\ov 6} {(1-\l)^{5/2}(1-4\l+\l^2)\ov (1+\l)^{9/2}}\, x_1^3 \bigg)  \, dx_2\wedge dx^+
\\
&  \qq  - {1+\l^2\ov \sqrt{(1-\l)(1+\l)^3}}\, x_1 \, dx_2\wedge dx^-\ .
\end{split}
\ee
Notice, that this expansion is invariant term by term under the transformation
\be 
\label{symm}
\l\to {1\ov \l} \ , \qq k\to -k \ ,\qq x^{\pm }\to \pm i x^\pm\ .
\ee
It turns out that this symmetry originates from that in \eqn{syymm}.

\section{Hamiltonian analysis} \label{Lagr}

In this section, we derive the Lagrangian governing the dynamics of the transverse degrees of freedom. This is obtained after fixing the light-cone gauge and imposing the Virasoro constraints. The analysis is general and holds 
for a generic model with an isometry.  Our analysis is similar to that of \cite{Callan:2004uv} with the essential, albeit crucial, difference that in that case the B-field was zero.

Our model has a metric of the form
\be 
ds^2 =  2 G_{+-} dx^+ dx^- + G_{++} (dx^+)^2 + g_{ij} dx_i dx_j 
\ee
and an antisymmetric tensor of the form
\be 
 B= b_{i+} dx_i \wedge dx^+ +  b_{i-} dx_i \wedge dx^- \ .
\ee
From \eqn{ppw} and \eqn{ppww}, we deduce that the various components have an expansion for large $k$ which is of the following type
\be
\begin{split}
& G_{+-}= 1 + {\cal O}(1/k)\ ,\quad G_{++}=  {\cal O}(1)\ + {\cal O}(1/k)\ ,
\\
& g_{11}= 1\ ,\quad g_{22} =  {\cal O}(1/k)\ ,
\\
& b_{1\pm}= 0\ ,
\\
& b_{2+} =  {\cal O}(1) + {\cal O}(1/k)\  ,\quad b_{2-} =   {\cal O}(1/k) \ ,
\end{split}
\ee
with the next corrections being of ${\cal O}(1/k^2)$. 
Note also that, since we will need the inverse metric the non-vanishing components are given by
$G^{--}=-{G_{++}\ov G_{+-}^2}$, $G^{+-}={1\ov G_{+-}}$ and $g^{ij}=(g^{-1})_{ij}$.

\no
Next we turn to the $\s$-model which, as we will see, it is convenient to be considered
first with a general world-sheet metric $\g_{ab}$. The action describing the $\s$-model is 
\be 
S = {1\ov 4\pi \a'} \int d^2\s\, \sqrt{-\g} \Big(\g^{ab} \del_a x^\m \del_b x^\n G_{\m\n} - \e^{ab} \del_a x^\m \del_b x^\n B_{\m\n}  + \a' R^{(\g)}\Phi\Big) \ ,
\ee
where the volume density is normalised as $\e^{01}=1/\sqrt{-\g}$ and consequently $\e_{01}=-\sqrt{-\g}$ . 
Note that in the conformal gauge with $\g_{ab} = \diag(1,-1)$ and by choosing $\a'=2$ we get \eqn{action}. 
Introducing now $h^{ab}$ by the following relation
\be
\label{hab}
h^{ab}= \sqrt{-\g} \g^{ab}\ , 
\ee
one can rewrite the action as 
\be 
\label{actga}
S = {1\ov 4\pi \a'} \int d^2\s \Big( h^{ab} \del_a x^\m \del_b x^\n G_{\m\n} - \bar \e^{ab} \del_a x^\m \del_b x^\n B_{\m\n} 
  + \a' \sqrt{-\g} R^{(\g)}\Phi\Big) \ ,
\ee
where here $\bar \e^{01}=1$. From this,
one can find the canonical momentum $\displaystyle p_\m = {\d S\ov \d \dot x^\m} $ which takes the form (we present it using the notation 
$\pi_\m = 2\pi \a' p_\m$)
\be
\begin{split}
& \pi_\m = h^{0a} G_{\m\n} \del_a x^\n -  B_{\m\n}x^{\prime \n}\ ,
\\
&  \dot x^\m = {\pi^\m \ov h^{00} }  - {1\ov h^{00}} \Big(h^{01}x^{\prime \m} -  B^\m{}_\n x^{\prime \n}\Big)\ ,
\end{split}
\ee
where, in the second equation, we have also included the inverse transformation of velocities in terms of momenta.  
The next step is to derive the Hamiltonian density $\cH= p_\m \dot x^\m - \cL$ which reads
\be\label{Ham}
\cH= {1\ov 2\pi \a'}  \bigg[{G^{\m\n}\ov 2 h^{00}}  \Big(\pi_\m  + B_{\m\a} x^{\prime \a}\Big)  \Big(\pi_\n  + B_{\n\b} x^{\prime \b}\Big) + 
{G_{\m\n}\ov 2 h^{00}} x^{\prime \m} x^{\prime \n}  - {h^{01}\ov h^{00}} \pi_\m x^{\prime \m}\bigg]  - { R^{(\g)}\Phi\ov 4\pi} \ .
\ee
Ignoring the dilaton, the Virasoro constraints follow from varying \eqref{Ham} with respect to $h^{01}$ and $h^{00}$. The result is
\be
\label{pppm}
\begin{split}
 & \pi_\m x^{\prime \m} = 0\ ,
 \\
 & G^{\m\n} \Big(\pi_\m  + B_{\m\a} x^{\prime \a}\Big)  \Big(\pi_\n  + B_{\n\b} x^{\prime \b}\Big) + G_{\m\n} x^{\prime \m} x^{\prime \n} =0\ .
\end{split}
\ee
In the conformal gauge  $\g_{ab} = \diag(1,-1)$ we have $h^{00}=1$ and $h^{01}=0$. Then these constraints reduce to the usual ones
$G_{\m\n} (\dot x^\m \dot x^\n +  x^{\prime \m} x^{\prime \n}) = 0 = G_{\m\n} \dot x^\m x^{\prime \n}$.

\no
However, in what follows, we will not choose the conformal gauge for the worldsheet metric. We will,  rather explicitly, write the above constraints for a generic worldsheet metric. The reason is that, since the background is curved it is not immediately obvious that one may choose the light cone gauge
\be
\label{lcg}
x^+= \tau \ .
\ee 
Nevertheless, we show that  this can be done by appropriately determining the world-sheet metric components $h^{00}$ and $h^{01}$, so that the variation of the action with respect to $x^-$  admits \eqn{lcg} as a solution
(this is in spirit similar to the approach in \cite{Callan:2004uv}). This is done in the Appendix \ref{AppB} and the end result is given by \eqn{answ}. In the light-cone gauge the constraints \eqn{pppm} 
assume the following form 
\be
\begin{split}
& \pi_- x^{\prime -} + \pi_i  x^{\prime i} =0 \ ,
\\
& G^{--} (\pi_- - b_{i-} x^{\prime i}) (\pi_- - b_{j-} x^{\prime j}) + 2 G^{+-}  (\pi_+ - b_{i+} x^{\prime i}) (\pi_-- b_{j-} x^{\prime j}) 
\\
& 
\phantom{xxxx} + g^{ij} (\pi_i + b_{i-} x^{\prime -}) (\pi_j + b_{j-} x^{\prime -})
+  g_{ij} x^{\prime i} x^{\prime j}= 0\ .
\end{split}
\ee 
The first constraint determines $x^-$ and the second one determines $\pi_+$ as
\be
\begin{split}
\pi_+ & =    b_{i+} x^{\prime i}
+ {G_{++}\ov 2 G_{+-}} (\pi_- - b_{i-} x^{\prime i})
\\
&  -{G_{+-}\ov 2( \pi_- - b_{k-} x^{\prime k}) } 
\Big(g_{ij} x^{\prime i} x^{\prime j}+ g^{ij} (\pi_i + b_{i-} x^{\prime -}) (\pi_j + b_{j-} x^{\prime -})\Big)\ .
\end{split}
\ee
 In this expression $x^{\prime -}$ should be replaced using the first constraint. 
 After doing so, we obtain that
 \be
\pi_+  =    b_{i+} x^{\prime i}
+ {G_{++}\ov 2 G_{+-}} (\pi_- - b_{i-} x^{\prime i})
 -{G_{+-}\ov 2( \pi_- - b_{k-} x^{\prime k}) } 
\Big(g_{ij} x^{\prime i} x^{\prime j}+ \tilde g^{ij} \pi_i \pi_j \Big)\ ,
\ee
where we have defined
\be
\tilde g^{ij} = \Big(\d^i_k - {x^{\prime i} b_{k-}\ov \pi_-}\Big) \,g^{k\ell} \Big(\d_\ell^j - {x^{\prime j} b_{\ell-}\ov \pi_-}\Big) \ .
\ee 
Note that this quantity is symmetric and that subsequently we will denote its inverse by $\tilde g_{ij}$.

\no
Our final goal is to find the Lagrangian associated to the physical degrees of freedom $x_1$ and $x_2$. 
The Lagrangian density corresponding to the Hamiltonian \eqref{Ham} is
$\cL = p_+ \dot x^+ + p_- \dot x^- + p_i \dot x^i - \cH$. In the light cone gauge and after implementing the constraints, so that $\cH=0$, we have that 
\be
\cL_{\rm l.c.} = p_i \dot x^i  + p_+\ .
 \ee
Note that, in the Lagrangian we have dropped the middle term, i.e. $p_- \dot x^- $  since upon partial integration in $\tau$ (in the action) it gives $-\dot p_-  x^-$ and 
 $\displaystyle \dot p_- = - {\del H\ov \del x^-}=0$.
The last equality holds since $x^-$  in our backgrounds is a cyclic coordinate. 
Furthermore, the light cone Hamiltonian density is $\cH_{\rm l.c.} =-p_+= -{\p_+\ov 2\pi \a'}$.  We now need 
to express the momenta $p_i$ in
 terms of the velocities $\dot x^i$. By using the Hamilton equation 
 \be
 \dot x^i =  {\del \cH_{\rm l.c.} \ov \del p_i} = -{1\ov 2\pi \a'}  {\del \pi_+ \ov \del p_i} 
 = -{\del \pi_+ \ov \del \pi_i} 
= {G_{+-}\ov  \pi_- - b_{k-} x^{\prime k} }  \tilde g^{ij} \pi_j \ ,
 \ee
 we obtain 
 \be
 \pi_i =  {\pi_- - b_{k-}x^{\prime k}\ov  G_{+-}}\,  \tilde g_{ij} \dot x^j\ .
 \ee
As a result, the Lagrangian density describing the evolution of the transverse degrees of freedom becomes
 \be\label{Lala}
 \begin{split}
2\pi \a' \cL_{\rm l.c.} = &\
 { \pi_- - b_{k-} x^{\prime k}\ov 2 G_{+-} } \tilde g_{ij} \dot x^i \dot x^j - {G_{+-}\ov 2( \pi_- - b_{k-} x^{\prime k}) } g_{ij} x^{\prime i}  x^{\prime j}
\\
 & +\Big( b_{i+}  - {G_{++}\ov 2 G_{+-}}b_{i-}\Big)x^{\prime i} + {G_{++}\ov 2 G_{+-}}\pi_-\ .
 \end{split}
 \ee 
 We should expand this expression up to ${\cal O}(1/k)$. In the expression above, we should in fact set $\pi_-=1$. The reason is that from \eqn{pppm}
 we have that $\p_-= G_{-+} h^{00} + b_{i-} x^{\prime i}$. Using \eqn{answ} for $h^{00}$ one sees that  $\pi_-$ simply becomes equal to unity.
 
 \subsection{Expanding the action}
 
 Using \eqref{Lala} one can obtain the following expansion for the light-cone action
 \be
 \cL_{\rm l.c.}  ={1\ov 2\pi \a' }\Big(  \cL^{(0)}_{\rm l.c.} +  {1\ov k} \cL^{(1)}_{\rm l.c.} +\dots \Big) + {\cal O}(1/k^2)   \ ,
 \ee
where 
\be
\label{lmiden}
\begin{split}
 \cL^{(0)}_{\rm l.c.}  = & \ha \Big(\dot x_1^2 + \dot x_2^2 - x_1^{\prime 2} - x_2^{\prime 2} \Big) 
-{1+\l^2\ov \sqrt{(1-\l)(1+\l)^3}} x_1 x^\prime_2 
\\
&-  {1\ov 8} \bigg[\Big({1-\l\ov 1+\l}\Big)^3 x_1^2 + {1+\l\ov 1-\l} x_2^2\bigg]
\end{split} 
\ee
and 
\be
\label{lena}
\begin{split}
 \cL^{(1)}_{\rm l.c.}  = & {1\ov 8} \bigg( \Big({1-\l\ov 1+\l}\Big)^3 x_1^2+ {1+\l\ov 1-\l} x_2^2\bigg)
(\dot x_1^2+x_1^{\prime 2})
\\
&
-{1\ov 8} \bigg( \Big({1-\l\ov 1+\l}\Big)^3 x_1^2 - {1+\l\ov 1-\l} x_2^2\bigg)\dot x_2^2 
+ {1\ov 8} \bigg( 3 \Big({1-\l\ov 1+\l}\Big)^3 x_1^2 +{1+\l\ov 1-\l} x_2^2\bigg) x_2^{\prime 2}\
\\
& +\bigg[{1\ov 24} {(1-\l)^{5/2} (1-16 \l + \l^2)\ov (1+\l)^{9/2}} x_1^2+{1\ov 8} {1+\l^2\ov \sqrt{(1-\l)^3 (1+\l)}} x_2^2
\bigg] x_1 x_2^\prime 
\\
& - \ha {1+\l^2\ov \sqrt{(1-\l)(1+\l)^3}} (\dot x_2^2 -\dot x_1^2 - x_1^{\prime 2}  -x_2^{\prime 2})x_1x_2^{\prime }
\\
&
 -{1+\l^2 \ov \sqrt{(1-\l)(1+\l)^3}} x_1 \dot x_1  x'_1 \dot x_2
\\
&-{1\ov 96} \bigg[ {(1-\l)^4(1+14 \l+\l^2)\ov (1+\l)^6}x_1^4 + \Big({1+\l\ov 1-\l}\Big)^2x_2^4\bigg]\ .
\end{split}
\ee
Notice that due to the light cone gauge this action is not manifestly Lorentz invariant. 
 
 The above expansion is not term by term invariant under the symmetry \eqn{symm} as it stands, the reason being that 
in the light cone gauge the coordinate $x^+$ has been set equal to $\tau$ (see \eqn{lcg}). However, the symmetry is recovered 
by reinstating $p^+$, which would have appeared if we had been using $x^+=p^+\tau$, instead of \eqn{lcg}.
From dimensional arguments $p^+$ counts as a derivative, either with respect to $\tau$ or $\s$.
Therefore, the simple derivative terms in \eqn{lmiden} 
and \eqn{lena} should be multiplied by $p^+$, the three derivative terms should be divided by $p^+$ and the
potential terms containing no derivatives should be multiplied by $(p^+)^2$. Then, \eqn{lmiden} 
and \eqn{lena} (after the inclusion of the overall $1/k$ factor) are separately invariant under
\be
\label{symnew}
 \l\to {1\ov \l}  \ ,\qq k\to -k\ ,\qq p^+\to i p^+\ .
\ee 

The interaction Lagrangian $ \cL^{(1)}_{\rm l.c.} $ leads to four particle amplitudes. Out of them only those corresponding to (elastic) scattering of two particles could be non-vanishing in an integrable theory and indeed this is the case. From these amplitudes we will compute the $S$-matrix.

\section{Computing the $S$-matrix }\label{secS-matrix}
In this section, we evaluate the S-matrix of the model and show that its form is consistent with integrability, namely that it allows only elastic scattering and that it obeys the Yang-Baxter equation.
\subsection{Spectrum of the model}
We first take advantage of the fact that by partial integration we may replace $x_1 x^\prime_2$ by $\ha (x_1 x^\prime_2-x_2 x^\prime_1)$ in \eqn{lmiden}. This is done for later convenience so that the matrix $M(p)$
below becomes Hermitian.
It is convenient to go to the momentum space by the use of
\be\label{mom}
x_i(\tau,\s)= {1\ov 2\pi} \int d^2 p\, e^{i (E \tau - p \s)} X_i(p)\ ,\quad X_i^*(p)= X_i(-p)\ ,\quad i=1,2\ ,
\ee
where we have defined the two-momenta $p^a = (E,p)$. In \eqref{mom} and in what follows $p$ will denote the spatial component of $p^a$. It is, then, straightforward to compute the leading term in the Lagrangian
\be
\label{lcmom}
\cL^{(0)}_{\rm l.c.}  =\ha  \int d^2 p X_i(-p) M(p)_{ij} X_j(p) = \ha  \int d^2 p X^\dagger(p) M(p) X(p)\ ,
\ee
where the two-dimensional square Hermitian matrix $M(p)$ is defined as 
\be
\label{massss}
\begin{split}
& M(p) =    \left( \begin{matrix} 
      E^2 - p^2 -m_1^2 & - i g p\\
        i g p & E^2 - p^2 -m_2^2 \\
   \end{matrix}
   \right)\ ,
   \\
   & 
   \phantom{x}
   m_1^2 = {1\ov 4}  \Big({1-\l\ov 1+\l}\Big)^3 \ ,\quad  m_2^2 = {1\ov 4} {1+\l\ov 1-\l} \ ,\quad g = -{1+\l^2\ov \sqrt{(1-\l)(1+\l)^3}}\ .
\end{split}
\ee
Note that $m_2^2\geqslant m_1^2$  for all values of $\l$ in the physical range $0\leqslant \l <1$. 
\no
The eigenvalues of the above matrix are given by the solutions of the quadratic equation for $M$, namely
\be
(E^2 - p^2 - m_1^2-M)(E^2- p^2-m_2^2-M) -  g^2 p^2 =0 \ .
\ee
These are given by
\be
M_{1,2}(p) = E^2-p^2 -{m_1^2+m_2^2\ov 2} \pm \sqrt{(m_2^2-m_1^2)^2/4 + g^2 p^2}\ ,
\ee
where the $+(-)$ goes with the index $1(2)$.\footnote{Note that when $m_1=m_2=\ha$ which happens for $\l=0$ the eigenvalues become simple, i.e. $M_{1,2}(p)= E^2 - (|p|\mp \ha)^2$.}
The corresponding eigenvectors are\footnote{The dispersion relation for the  particle $Y_i$ is found by setting the eigenvalue $M_i=0$ from which we may
express $E$ in terms of $p$. Then on-shell the parameter $\a$ in \eqn{ab} takes the form $\a(Y_1)\equiv \a_1 = p^2 -E^2 +m_2^2$ and
$\a(Y_2)\equiv \a_2 = E^2 -p^2 -m_1^2$.
} 
\be
\label{ab}
\begin{split}
& Y_1= {1\ov \sqrt{\a^2-\b^2}} \left( \begin{matrix}
\a
\\
\b
\end{matrix}
\right)\ ,\qquad 
 Y_2= {1\ov \sqrt{\a^2-\b^2}}  \left( \begin{matrix}
 \b
  \\
\a
\end{matrix}
\right)\ ,
\\
&
\a = (m_2^2-m_1^2)/ 2 + \sqrt{ (m_2^2-m_1^2)^2/4  + g^2p^2}\ ,\qquad \b =  i g p\ .
\end{split}
\ee
Then we define the Unitary matrix
\be 
U (p)=  {1\ov \sqrt{\a^2-\b^2}} \left( \begin{matrix}       \a & -\b \\
      -\b & \a \\
   \end{matrix}\right)\ .
\ee 
It obviously holds that $M = U^\dagger M_{ \diag} U$, where $ M_{\rm \diag} = \diag(M_1,M_2)$. Furthermore, note that  the relations $U^\dagger(p)=U(-p)= U^*(p)$ and 
$X_i(-p)= X_i^*(p)$ are also valid. 
One can then pass to the new basis 
\be
\label{newba}
Y(p)=U(p)X(p)\ ,\qq  X(p)= U^\dagger(p) Y(p)\ ,
\ee
in which the Lagrangian density \eqn{lcmom} becomes
\be
\begin{split}
\label{lcmom}
\cL^{(0)}_{\rm l.c.}  &=\ha  \int d^2 p\, Y^\dagger(p) M_{\diag }(p) Y(p) 
\\
& =  \ha  \int d^2 p\, \Big(M_1(p) Y_1(-p) Y_1(p) + M_2(p) Y_2(-p)Y_2(p)\Big)\ .
\end{split}
\ee
Similarly to the case with the $X_i$'s we also have that  $Y_i(p)^*=Y_i(-p)$.  The on-shell condition for the particles represented by the fields 
$Y_{1,2}$ are found by setting $M_{1,2}=0$, respectively. Explicitly, we have
\be
\label{onshell}
\begin{split}
Y_1: &  \qq E= \sqrt{ p^2 +{m_1^2+m_2^2\ov 2} - \sqrt{(m_2^2-m_1^2)^2/4 + g^2 p^2}}\ ,
\\
Y_2: &  \qq E= \sqrt{ p^2 +{m_1^2+m_2^2\ov 2} + \sqrt{(m_2^2-m_1^2)^2/4 + g^2 p^2}}\ ,
\end{split}
\ee
where $m_1$, $m_2$ and the coupling $g$ are given by \eqn{massss}. 
One may check, using the specific expression, 
 that $E$ for $Y_1$ is real for all values of the spatial momentum $p$ and the parameter $\l$.
Subsequently,  we will call particle $1 (2)$ the one corresponding to the field $Y_1 (Y_2)$. It is these particles which will scatter and for which the requirements of integrability will hold. 

\begin{figure}
\centering
\includegraphics[width=0.45\textwidth]{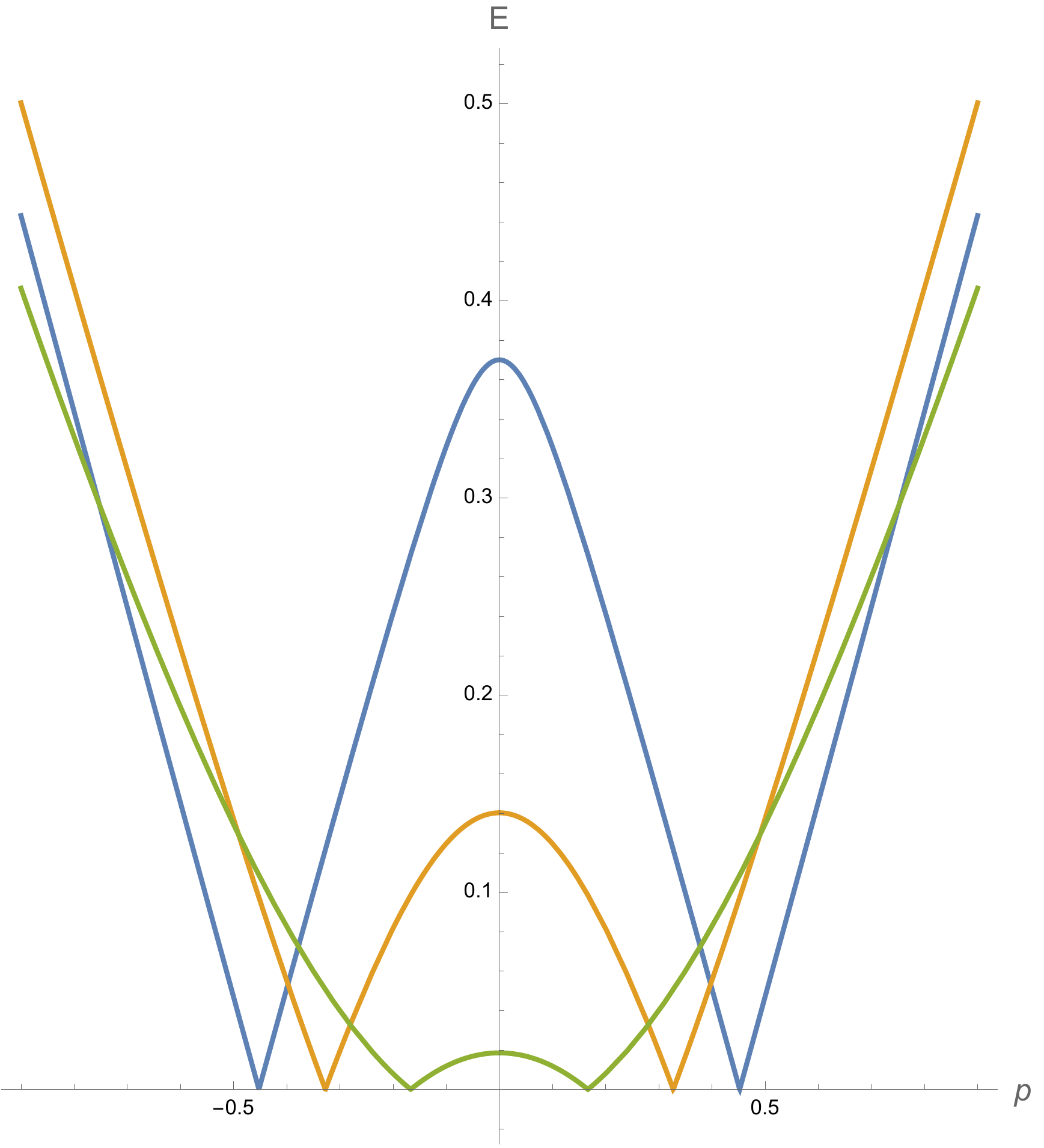}\hskip 1 cm 
\includegraphics[width=0.45\textwidth]{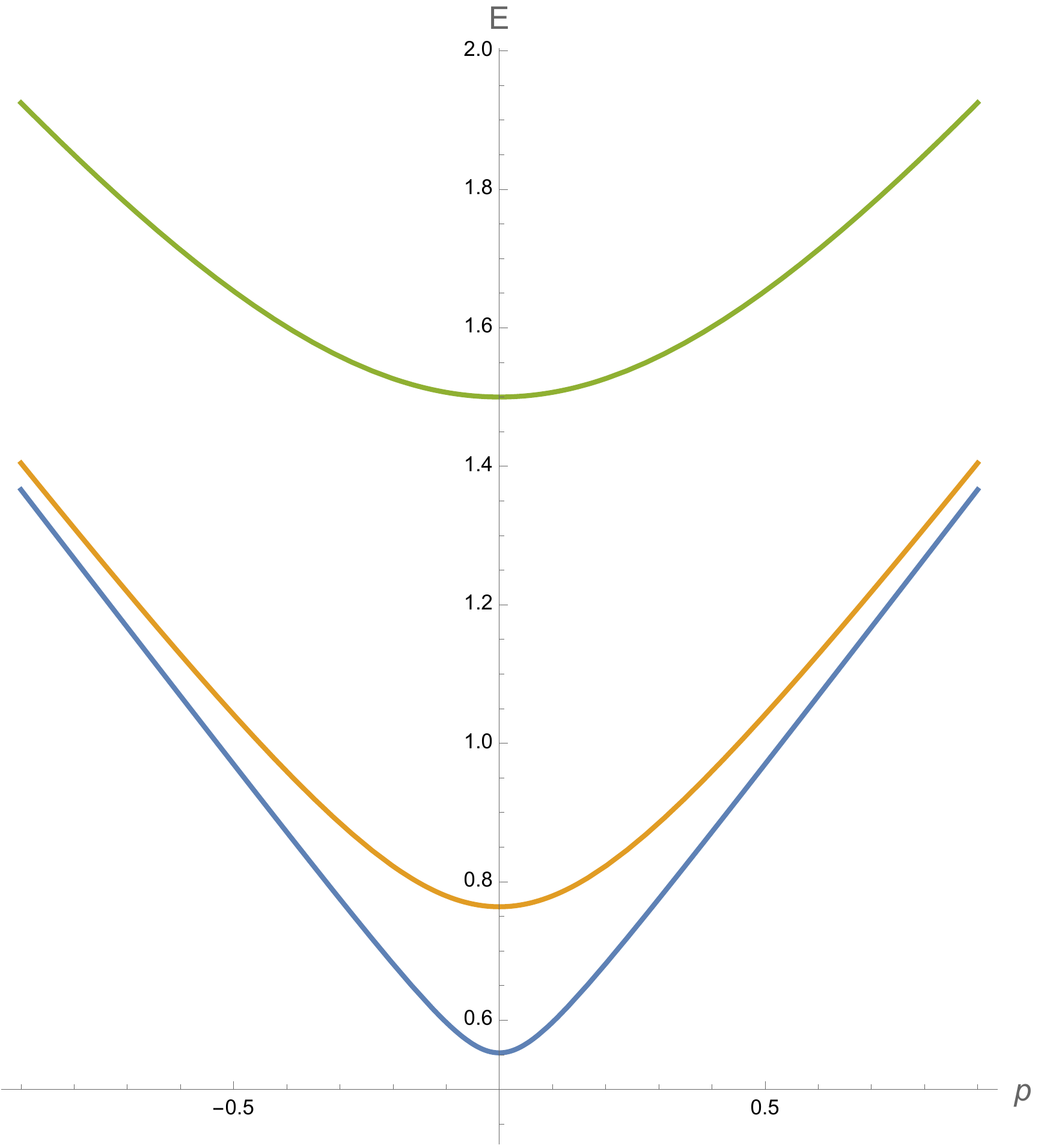}
\vskip -.25 cm
\caption{The dispersion relations for particle 1 (left) and particle 2 (right) found using \eqn{onshell}. 
The colors correspond to different values of $\l$: For $\l=0.1$ (blue), $\l=0.4$ (brown) and $\l=0.8$ (green).}
\label{Fig.1}
\end{figure}

The above dispersion relations are depicted in Fig. \ref{Fig.1}.
for various non-zero values of $\l$. The dispersion relation of particle 1 is the most interesting one. 
In that case $E$ vanishes for the values of the spatial momentum $\displaystyle p= \pm \ha \sqrt{1-\l\ov 1+\l}$. At these values the first derivative of the dispersion relation becomes discontinuous. At $p=0$ we have that $E=m_1$ which is a local maximum. 
The dispersion relation for particle 2 has just a minimum at $p=0$ given by $E =m_2$.
For the special case $\l=0$ the two dispersion relations greatly simplify and read 
\be
\label{onshell-1}
\begin{split}
Y_1: &  \qq E= \Big||p|-\ha \Big |\ ,
\\
Y_2: &  \qq E =  \Big||p|+\ha \Big |\ .
\end{split}
\ee
In this case, for particle 1 the derivative of the dispersion relation becomes discontinuous for $p=0$ as well as for $p=\pm \ha$.  Similarly, the derivative of the dispersion relation for particle 2 becomes discontinuous for $p=0$.

\no
\subsection{Interaction Lagrangian and the S-matrix}\label{Lint}
Next we consider the interaction term $\cL^{(1)}_{\rm l.c.}$ which in momentum space takes the form
\be
\label{l1lc}
\begin{split}
{\cal L}_{\rm l.c.}^{(1)} = & \int {d^2p_1\cdots d^2 p_4\ov (2\pi)^3}\,  \d^{(2)}(p_1+\cdots +p_4)
\\
& \Big({\cal J}_1\, X_1(p_1) X_1(p_2)X_1(p_3)X_1(p_4)+{\cal J}_2\, X_1(p_1) X_1(p_2)X_2(p_3)X_2(p_4)
 \\
& +{\cal J}_3\, X_2(p_1) X_2(p_2)X_2(p_3)X_2(p_4)+{\cal J}_4\, X_1(p_1) X_1(p_2)X_1(p_3)X_2(p_4)
 \\
&  + {\cal J}_4\, X_1(p_1) X_2(p_2)X_2(p_3)X_2(p_4)\Big)\ ,
\end{split}
\ee
where
\begin{eqnarray}\label{Js}
&&{\cal J}_1=-{1 \ov 8}\Big({1-\l\ov 1+\l}\Big)^3 \big(E_1 E_2+ p_1 p_2\big)
-{1 \ov 96}{(1-\l)^4 (1+14 \l +\l^2)\ov (1+\l)^6}\ ,
 \nonumber \\
&&{\cal J}_2=-{1 \ov 8}{1+\l\ov 1-\l}\big(E_1 E_2 + p_1 p_2\big)+{1 \ov 8}\Big({1-\l\ov 1+\l}\Big)^3\, \big(E_3 E_4 
-3 \, p_3 p_4\big) \ ,
\nonumber \\
&&{\cal J}_3=-{1 \ov 8}{1+\l\ov 1-\l}\big(E_3 E_4+p_3 p_4\big)-{1 \ov 96}\Big({1+\l\ov 1-\l}\Big)^2
\ ,
\nonumber \\
&&{\cal J}_4=-{i \ov 24}{(1-\l)^{5/2} (1-16 \l +\l^2)\ov (1+\l)^{9/2}} p_4
\\
&&\qq\quad +{i \ov2 }{1+\l^2 \ov \sqrt{(1-\l)(1+\l)^3}}\big( p_2 p_3 p_4 + E_2 E_3 p_4 - 2 E_2 p_3 E_4 \big)\ ,
\nonumber \\
&&{\cal J}_5=-{i \ov 8}{1+\l^2 \ov \sqrt{(1+\l)(1-\l)^3}} p_4
+{i \ov 2} {1+\l^2 \ov \sqrt{(1-\l)(1+\l)^3}}\big( p_2 p_3 p_4 - E_2 E_3 p_4\big)\, .
\nonumber
\end{eqnarray}
Subsequently, we have to perform the basis change \eqn{newba} and consider specific processes in which four particles 
are involved. 
 Their two-momenta are given by
\be
p_i^a = \big( E_i,p_i\big)\, ,\quad i=1,2,3,4\ .
\ee 
The on-shell condition constrains
$E_i$ to be given by the first of  \eqn{onshell} if it is particle 1, or by the
second one if it is particle 2. 
 \no
 
 First consider amplitudes corresponding to particles transmutation. These are the processes  
 \be\label{2to2}
 1+1\to 1+2\,, \quad  1+1\to 2+2\,, \quad   2+2\to 2+1\, ,
 \ee
 as well as the ones with reverse arrows. 
All of these process are allowed by the kinematic constraints imposed by the $\d$-function 
constraints in \eqn{l1lc}\footnote{The $\d$-function constraint in \eqn{l1lc} treats all external particles as incoming.
In order to comply with this the conservation laws  should be written as $\sum_i E_i = 0 = \sum_i p_i$. For the dispersion relations of the outgoing particles we should use still the ones in \eqn{onshell} but with a negative overall sign. This flip of signs has already been taken into account in \eqn{fuh} and \eqn{fuh1} below. Nevertheless, when it comes to extracting the amplitudes we fully conform with  \eqn{l1lc}.  
\label{ff}}
\be
\label{fuh}
E_1 + E_2 = E_3 + E_4\ ,\qq p_1 + p_2 = p_3 + p_4\ ,
\ee
which allow for two independent components of spatial momenta. 
 In addition, we have processes corresponding to particle creation 
 \be
 \label{check}
1 \to  1+1+1 \ ,\quad 2\to 2+1+1 \, ,\quad 2\to 1+1+1  \, ,
\ee
and 
\be
\label{uncheck}
\begin{split}
& 1\to 1+1+2 \, ,\quad 1\to 1+2+2 \, ,\quad 1\to 2+2+2 \, , 
\\
&  2\to 2+2+1 \, , \quad  \, 2\to 2+2+2\ ,
\end{split}
\ee
as well as the ones with $1$ and $2$ interchanged. In addition, we may have particle destruction corresponding
to reversing the arrows above.  However, as far as kinematics is concerned, these are not any different.
The kinematic constraints for these cases are given by
\be
\label{fuh1}
E_1 = E_2 +E_3 + E_4\ ,\qq p_1 = p_2 + p_3 + p_4\ .
\ee
We have verified that for the three ones appearing in \eqref{check} the above constraints can be satisfied in certain 
ranges of $\l$ and of the spacial momenta. For the remaining decaying processes of \eqref{uncheck}, we were unable to do so and we believe that it is not 
possible for the following reason. From \eqn{onshell} we see that the particle of type 2, for generic values of $\l$ and of its spatial momentum, is more energetic than the type 1 particle. Hence, it seems impossible  to produce a particle (let alone more) of type 2 from a single particle of type 1. Similarly it seems impossible to produce more that one  type 2 particles having started from s single particle of   type 2. 

\no
Even if allowed by kinematics,
 {\it all} of the above processes in \eqref{2to2} and \eqref{check} have vanishing amplitudes. We have checked this numerically. In fact, for this to happen the form of the interaction vertices and their dependence on the 
$\l$-parameter is crucial. Indeed, even the slightest modification in the form of the functions appearing in \eqref{Js}  leads to non-vanishing amplitudes.  Of course, the particular functional dependence of these functions on the deformation parameter $\l$ and on the momenta is also crucial for the model to be integrable, as well. 
We, thus, reach the important conclusion that our theory does not support neither flavour-changing processes \eqref{2to2} nor particle creation \eqref{check}. 
In addition, due to the absence of terms of ${\cal O}(1/k^{1/2})$ and ${\cal O}(1/k^{3/2})$ the three- and five-point {\it contact} amplitudes that would have been originating from such parts of the Lagrangian are zero. 
Consequently, {\it all}  three- and five-point amplitudes are zero in our model. This observation combined with the results of section \ref{Lint} implies the absence of particle production up to ${\cal O}(1/k^{3/2})$.
These are strong hints that the theory after the pp-wave limit preserves the integrability of the parent theory.

 \no
 We now turn to the processes which integrability allows to have non-vanishing amplitudes. These can be either two-particle scaterring of the same kind 
 \be
 \label{111klp}
 1_{p_1}+1_{p_2}\to 1_{p_3}+1_{p_4}\  , \qquad 2_{p_1}+2_{p_2}\to 2_{p_3}+2_{p_4}\, ,
 \ee 
 as well as of a different kind 
  \be
 \label{111klpe}
 1_{p_1}+2_{p_2}\to 1_{p_3}+2_{p_4}\, .
 \ee 
 For scattering of two particles of the same kind conservation of momentum and energy \eqn{fuh} force that either
\be
\label{first}
p_1^a = \big( E_1,p_1\big)\, ,\quad  p_2^a = \big( E_2,p_2\big)\, , \quad p^a_3 =- p^a_1\, \ ,\quad  p^a_4 =-p^a_2 \, .
\ee
or that 
\be
\label{second}
p_1^a = \big( E_1,p_1\big)\, ,\quad  p_2^a = \big( E_2,p_2\big)\, , \quad p^a_3 =- p^a_2\, \ ,\quad  p^a_4 =- p^a_1 \, ,
\ee
where the negative sign in the above identification of momenta has been explained in footnote \ref{ff}.
The corresponding amplitudes will be given below. 

\no
 For scattering of two particles of a different kind conservation of momentum and energy is guaranteed when \eqn{first} is satisfied. 
 However, there are solutions of the kinematic conditions different than \eqn{first} when $|p_{1,2}|\leqslant {1\ov 2}\sqrt{{1-\l \ov 1+\l}}$.
 We have checked numerically that for 
 these the corresponding amplitude vanishes. \\
 For $|p_{1,2}|\geqslant {1 \ov 2}\sqrt{{1-\l \ov 1+\l}}$ the 
only solution  we were able to find is that in \eqn{second} corresponding to exchange of particle momenta. The fact that this is only allowed when the magnitudes of the momenta obey the above lower bound condition is interesting since it is related to Fig. \ref{Fig.1}. Hence, we present some steps. 
The on-shell \eqn{onshell} conditions can be written in the form $E=\sqrt{a\pm b}$, 
with the obvious definitions for $a$ and $b$. 
Using the specific forms for the various parameters in terms of $p$ and $\l$ we may easily check that 
$b^2+(a-g^2/2)^2= a^2$. Then, the conservation of energy is $\sqrt{a_1+b_1}- \sqrt{a_1-b_1}= \sqrt{a_2+b_2}
-\sqrt{a_2-b_2}$, where the subscripts indicate the use of momentum $p_{1,2}$ in the definition of $a$ and $b$. 
Squaring the previous relation we get that 
\be 
a_2-a_1= \Big |a_2-{g^2\ov 2} \Big | - \Big|a_1-{g^2\ov 2} \Big |\ .
\ee
Hence for this to hold the conditions $a_i\geqslant g^2/2$, $i=1,2$ should be obeyed. This is equivalent to the 
condition on the magnitudes of the momenta we stated above. 
However, even though it is allowed kinematically, we have checked numerically that the corresponding amplitude is zero due to the dynamics of the theory. In conclusion,
\be\label{12to21}
{\rm Amp}(1_{p_1}+2_{p_2}\to 1_{-p_2}+2_{-p_1})  =0 \quad \Longrightarrow\quad   S_{12}^{21}(p_1,p_2)=0\, .
\ee

In order to proceed computing the non-vanishing amplitudes we define for convenience the functions 
\be
 f = {1 - \l\ov 1 + \l}\  \qq g_1 = {1 + 14 \l +  \l^2\ov (1 + \l)^2}\ ,\qq  g_2 = {1 -16 \l +  \l^2\ov (1 + \l)^2}\ ,
\ee
as well as
\be 
 \label{abc...}
 \begin{split}
 & A_{12} =    -{1\ov 4 f^2} + {1\ov 2 f}\big( E_1^2  + p_1^2 + E_2^2 + p_2^2\big)\ ,
\\
 &  B_{12}= 2  \Big((f^3-f^{-1})  E_1 E_2  - (3 f^3+f^{-1})  p_1  p_2\Big)\ ,
\\
 &  C_{12} = -{f^3\ov 4} \bigg( g_1 f-2 \Big(   E_1^2 + p_1^2 +  E_2^2 + p_2^2\Big)   \bigg)\ ,
 \\
  &  D_{12}= {i  g\ov 2}   \bigg(8 f\, E_1 E_2 p_2 - {1\ov f} p_1 + 4 p_1\big(E_2^2 - 3 p_2^2\big)\bigg)\ ,
 \\
 & E_{12} = \ha \biggl( f^3 \big(E_2^2 - 3 p_2^2\big) -{1\ov f} \big( E_1^2 + p_1^2\big) \biggl) \ ,
 \\
 & F_{12}= - {i\ov 2}  \bigg( 8 g\, E_1 p_1 E_2 + f^{5/2} g_2 p_2 - 4 g\, p_2\big(E_1^2+ p_1^2 \big) \bigg) \ .
 \end{split}
 \ee
 The indices $1,2$ above indicate that the various functions depend on the components of the two particles $p_1$ 
 and $p_2$. 
 The expressions for $A_{12}$,  $B_{12}$  and $C_{12}$ are invariant under the interchange of $p_1$ and $p_2$, e.g. $A_{12}=A_{21}$,
 whereas this is not the case for $D_{12}$,  $E_{12}$  and $F_{12}$.  
 
 \no
 All three amplitudes corresponding to
 \eqn{111klp} can be expressed in terms of the functions defined in \eqn{abc...}.
 Indeed, we have that 
 \be
 \label{ampl1}
  {\rm Amp}(1+1\to 1+1)  = {1\ov k} {i\ov (2\pi)^3} { \cA(1+1\to 1+1)\ov  (\a_1^2-\b_1^2) (\a_2^2-\b_2^2)}\ ,
 \ee
 where
\be 
 \begin{split}
 & \cA(1+1\to 1+1)  =    \b_1^2  \b_2^2\, A_{12}
+  \a_1\a_2 \b_1\b_2\, B_{12} +  \a_1^2 \a_2^2 \, C_{12} 
\\
& \ + \a_1 \b_1\b_2^2\, D_{12} +  \a_1^2 \b_2^2\, E_{12} +  \a_1^2\a_2 \b_2\, F_{12}  
+ \a_2 \b_2\b_1^2\, D_{21} +  \a_2^2 \b_1^2\, E_{21} +  \a_2^2\a_1 \b_1\, F_{21}  \ 
 \end{split}
 \ee
 and also (on-shell)
 \be
 \a_i = p_i^2 -E_i^2 +m_2^2\, ,\quad  \b_i = i g  p_i\ ,\quad i=1,2\ .
 \ee
 %
 Similarly, we have obtained
  \be
   \label{ampl2}
  {\rm Amp}(1+2\to 1+2)  = {1\ov k} {i\ov (2\pi)^3} { \cA(1+2\to 1+2)\ov  (\a_1^2-\b_1^2) (\a_2^2-\b_2^2)}\ ,
 \ee
 where
\be 
 \begin{split}
 & \cA(1+2\to 1+2)  =    -\b_1^2  \b_2^2\, E_{21}
-  \a_1\a_2 \b_1\b_2\, B_{12} -  \a_1^2 \a_2^2\,  E_{12} 
\\
& \ - \a_1 \b_1\b_2^2\, F_{21} -  \a_1^2 \b_2^2\, C_{12} -  \a_1^2\a_2 \b_2\, F_{12}  
- \a_2 \b_2\b_1^2\, D_{21} -  \a_2^2 \b_1^2\, A_{12} -  \a_2^2\a_1 \b_1\, D_{12}  \ 
 \end{split}
 \ee
  and also (on-shell)
 \be
 \a_1 = p_1^2 -E_1^2 +m_2^2\, ,\quad a_2 = E_2^2 -p_2^2 -m_1^2\, ,\quad   \b_i = i g  p_i\ ,\quad i=1,2\ .
 \ee
 Finally,      
  \be
   \label{ampl3}
  {\rm Amp}(2+2\to 2+2)  ={1\ov k} {i\ov (2\pi)^3} { \cA(2+2\to 2+2)\ov  (\a_1^2-\b_1^2) (\a_2^2-\b_2^2)}\ ,
 \ee
 where
\be 
 \begin{split}
 & \cA(2+2\to 2+2)  =    \b_1^2  \b_2^2\, C_{12}
+ \a_1\a_2 \b_1\b_2\, B_{12} +  \a_1^2 \a_2^2\,  A_{12} 
\\
& \ +\a_1 \b_1\b_2^2\, F_{21} +  \a_1^2 \b_2^2\, E_{21} +  \a_1^2\a_2 \b_2\, D_{21}  
+ \a_2 \b_2\b_1^2\, F_{12} +  \a_2^2 \b_1^2\, E_{12} +  \a_2^2\a_1 \b_1\, D_{12}  \ 
 \end{split}
 \ee
 and also (on-shell)
 \be
 \a_i = E_i^2 -p_i^2 -m_1^2\, ,\quad  \b_i = i g  p_i\ ,\quad i=1,2\ .
 \ee
 Note that the amplitudes  $  {\rm Amp}(1+1\to 1+1) $ and  $  {\rm Amp}(2+2\to 2+2)$ are invariant under the interchange $p_1$ and $p_2$ as they should. 
This is of course not the case for the amplitude $  {\rm Amp}(1+2\to 1+2) $. The symmetry \eqn{symnew} is also a symmetry of 
 the amplitudes. To see this we should as in the case of the light cone action, discussed below \eqn{lena}, 
 reinstate the parameter $p^+$. To do so we just have to insert in the various terms in \eqn{abc...} appropriate powers of $p^+$ so that the functions $A_{12}, B_{12},\dots, F_{12}$ have units of square momentum.  Then these functions under the transformation \eqn{symnew} pick up an overall minus sign which is compensated by the flip of the sign of the overall factor $1/k$ in \eqn{ampl1}, \eqn{ampl2} and \eqn{ampl3}.
 
 Finally, note that all the expressions \eqn{ampl1}, \eqn{ampl2} and \eqn{ampl3} are purely imaginary. This implies that the $S$-matrix is of the form $\displaystyle S= {\mathbb 1}+{i\ov k} (\cdots)+{\cal O}(1/k^2)$ and indeed to the given order is unitary, i.e. obeys $S^\dagger S= {\mathbb 1} + {\cal O}(1/k^2)$.  It would be certainly interesting to calculate the leading correction to the S-matrix that we have just computed.
 Next we write the matrix elements of the S-matrix in terms of the amplitudes we have calculated
 \ba
 \label{S-matrix}
 &&S_{11}^{11}(p_1,p_2)=1+{\rm Amp}(1+1\to 1+1), \quad S_{22}^{22}(p_1,p_2)=1+{\rm Amp}(2+2\to 2+2)\, , 
 \nonumber
 \\
 &&S_{12}^{12}(p_1,p_2)=1+{\rm Amp}(1+2\to 1+2),\quad  S_{21}^{21}(p_1,p_2)=1+{\rm \tilde Amp}(1+2\to 1+2)\, ,
 \nonumber
 \\
 &&S_{f_1f_2}^{f_1'f_2'}(p_1,p_2)=0, \,\,\,\,\,\,{\rm for \,\,all\,\, other\,\, choices\,\, of\,\,} \{f_1,f_2,f_1',f_2'\}\, ,
 \ea
 where ${\rm \tilde Amp}(1+2\to 1+2)$ is ${\rm Amp}(1+2\to 1+2)$ but with $p_1$ and $p_2$ exchanged, i.e. $p_1\leftrightarrow p_2$. Notice that there is nothing special for all the above results when $\l=0$.
 
 A last comment is in order. One can easily verify that our S-matrix satisfies the Yang--Baxter equation\footnote{All flavour indices in \eqref{YB} take values 1 or 2 according to the particle they represent and which takes part in the interaction vertex.}
 \be\label{YB}
 S_{bc}^{f_2' f_1'}(p_2,p_3)S_{af_3}^{f_3' c}(p_1,p_3)S_{f_1f_2}^{ab}(p_1,p_2)=
S_{\tilde c \tilde a}^{f_3' f_2'}(p_1,p_2)S_{f_1 \tilde b}^{ \tilde c f_1'}(p_1,p_3)S_{f_2f_3}^{\tilde a\tilde b}(p_2,p_3) \ ,
 \ee
 as long as the amplitude \eqref{12to21} is zero, or more precisely, as long as the S-matrix assumes the form given in equations \eqref{S-matrix}. This statement holds to all orders in the large $k$ expansion and its validity should be verified order by order in perturbation theory.

 \section{Concluding remarks}\label{conclusions}
 
Classical integrability does not automatically imply that the corresponding quantum theory will be integrable, too. This is so because conservation laws of the higher conserved charges may suffer from quantum anomalies \cite{Abdalla:1,Abdalla:2}. One way to argue that the quantum theory is integrable is to show that its scattering S-matrix factorises and does not support particle production. In certain 2-dimensional integrable models, it is often the case that the S-matrix can be found exactly. However, this is not the case for a generic  integrable theory.  
One may try to evaluate the S-matrix by expanding around the trivial flat space vacuum. However in such an approach, the connection to integrability is lost because the spectrum contains massless excitations.

It was the aim of our work  to initiate the perturbative study of the S-matrix of  generic integrable $\s$-models and provide an independent way to confirm its integrability. 
In particular, we develop a method which re-establishes the link between integrability and factorisation of the S-matrix. This is accomplished by expanding around a vacuum which supports massive excitations. Our method can be applied to any integrable theory with one or more isometries. The strategy is to add to the $\sigma$-model action, assumed of Euclidean signature, a timelike spectator field $t$. 
One then takes the Penrose limit of the geometry around a null geodesic involving the isometry of the initial background and $t$ and systematically expands order by order around this geodesic. Subsequently, one fixes the light-cone gauge and imposes the Virasoro constraints so that he is left only with the transverse degrees of freedom. All transverse excitations are massive and the properties of factorisation and no particle production can now be unambiguously addressed. 

We have exemplified our method in the case of the isotropic $\l$-model with group $SU(2)$. We have calculated the pp-wave background of the aforementioned model, as well as the post pp-wave corrections which account for the interactions of its two massive excitations. We have presented the Lagrangian governing the propagation of the physical modes up to ${\cal O}(1/k^{3/2})$. It inherits a certain non-perturbative symmetry which is present in the original $\l$-model. Subsequently, we have calculated the S-matrix  and have shown that it does not exhibit particle production up to ${\cal O}(1/k^{3/2})$. Furthermore, we showed that it satisfies the Yang-Baxter equation up to the same order. Thus, the structure and properties of the S-matrix are consistent with integrability.

A word of caution regarding the equivalence between the initial integrable theory and the one obtained after expanding its action around the aforementioned null geodesic is in order. We do not have a general proof that the integrability of the parent theory will be inherited to the final theory.
In the process, we have fixed the light-cone gauge and have solved  the equation of motion for one of the light-cone coordinates. Most importantly, we have imposed the Virasoro constraints which amounts to projecting out some states from the Hilbert space of the parent theory. However, our expectation is that the gauge-fixed theory will retain  the integrability of the original model at the quantum, as much as at the classical level. This does not imply that the imposition of the Virasoro constraints doesn't affect certain properties of the theory. 
In particular, it will be interesting to compute the $\beta$-function of the action of the reduced theory since, given the reduction of the 
degrees of freedom, it is not obvious that it will stay the same. Indeed, as an analogy, 
we mention that the $\beta$-functions for the $SU(2)$ and the $SU(2)/U(1)$ $\l$-models are drastically  different \cite{Itsios:2014lca}, 
as the projection procedure to the coset changes also in that case the degrees of freedom of the model.

In this work we have shown that, in accordance with integrability, the decay of a particle to give three particles, or the fusion of three particles to give one are not possible processes. It would be very interesting to extend our calculation for the $\l$-model up to 
${\cal O}(1/k^2)$ and show that the amplitudes in which two particles scatter to produce four are also zero. This calculation would also enable us to compute the leading correction to the S-matrix computed in the present paper. Notice that this calculation would require to take into account the  ${1/ k}$ quantum corrections to the action of the $\l$-model and will unable us to further check the Yang--Baxter equation to 
a higher order in perturbation theory and also check if the six-point amplitudes factorise.
Another interesting direction would be to apply the method of this paper to $\l$-models based on more generic groups and most importantly to the generalisation of the $\l$-models \cite{Georgiou:2017jfi,Georgiou:2018hpd,Georgiou:2018gpe} that interpolate between two different
 CFTs in the IR and in the UV.

\no 
Furthermore, our method could also be applied to other classes of integrable models,
 such as the Yang-Baxter and bi-Yang-Baxter $\s$-models \cite{Klimcik:2002zj,Klimcik:2008eq}.
In addition, we mention that the method presented in this paper, may provide a useful tool in determining the quantum corrections of
 classically integrable $\s$-model models, as well as the tensor structure and the symmetries of their S-matrix. 

\subsection*{Acknowledgements}

We would like to thank K. Siampos for a thorough reading of the manuscript.
This research work was supported by the Hellenic Foundation for Research and Innovation (H.F.R.I.) under the ``First Call for H.F.R.I. Research Projects to support Faculty members and Researchers and the procurement of high-cost research equipment grant'' (MIS 1857, Project Number: 16519).

 \appendix
 
 \section{Various plane-wave limits}\label{AppA}
 
 In this appendix, we consider various limits giving rise to plane-wave backgrounds originating from \eqn{g.su2} and \eqn{B.su2}. Unlike the main text and for notational convenience, we will use $u$ and $v$ for the light 
 cone coordinates instead of $x^+$ and $x^-$.
 
\subsection{The  $ \l\to 1$ limit of the plane-wave background \eqn{ppw} }

Consider the plane-wave background \eqn{ppw} in the limit that parameter $\l\to 1$. To make sense of this limit 
first let the redefinition
\be 
x_+ =  y_+ \sqrt{2(1-\l)}\, ,\quad  x_- ={ y_- \ov \sqrt{2(1-\l)}}  
\ee
and then take the limit $\l\to 1$. We find the Lagrangian density
\be
\begin{split}
\cL= \,& \del_+ y^+ \del_- y^- +  \del_+ y^- \del_- y^+  + \del_+ x_1\del_- x_1+ \del_+ x_2\del_- x_2  -  x_2^2 \del_+ y^+ \del_- y^+ 
\\
& -x_1(\del_+ x_2 \del_- y^+ - \del_+ y^+ \del_- x_2)\ .
\end{split}
\ee
This limit resembles a bit the non-Abelian T-duality limit on the background \eqn{g.su2} and \eqn{B.su2} in the sense that $\l$ in both
cases is taken to unity. However, this is only an apparent similarity since in the non-Abelian limit, the limit is taken in conjunction with the $k\to \infty$ limit \cite{Sfetsos:2013wia} whereas here the latter limit in taken first and independently.

\subsection{The plane-wave limit of the non-Abelian T-dual of $S^3$}

Letting $\a=r/( 2k)$ and $k\to \infty$ in the background \eqn{ppw} we get the non-Abelian T-dual for the PCM for $SU(2)$ with
\be
\begin{split}
&ds^2 = L^2\Big(-dt^2 + dr^2 + { r^2\ov 1+r^2} \big(d\beta^2+\sin^2\beta\, d\gamma^2\big)\Big)\ ,
\\
&
 B =- L^2 {r^3\ov 1+r^2} \sin\beta\,\text{d}\beta\wedge\text{d}\gamma\ ,
\end{split}
\ee
where we have multiplied with an arbitrary factor $L$ and added a time coordinate. 
Let 
\be
r=u\ , \qq t = u-{v\ov L^2} \ ,\qq \b= {\r  \ov L}\ ,\qq L\to \infty\ .
\ee
Note that, unlike the main text and for notational convenience, we will use $u$ and $v$ for the light 
 cone coordinates instead of $x^+$ and $x^-$.
We obtain that 
\be
ds^ 2= 2 du dv + {u^2\ov 1+u^2} (dx_1^2 + dx_2^2)\ ,\qq B = - {u^3\ov 1+u^2} dx_1\wedge dx_2\ ,
\ee 
where we have passed from polar $(\r,\g)$ to Cartesian coordinates $(x_1,x_2)$.
This is a plane wave in its Rosen form . We may brink it in its Brinkmann form by performing a suitable coordinate transformation.
Sparing the details we find that
\be
\begin{split}
& ds^2 = 2 du dv + dx_1^2 + dx_2^2 -3 { x_1^2 + x_2^2 \ov (1+u^2)^2}\, du^2 \ ,
\\
& H=dB = -{3+u^2\ov 1+u^2}\, du\wedge  dx_1\wedge dx_2\ .
\end{split}
\ee
Judging from the results the plane-wave of the non-Abelian T-dual for the $SU(2)$ PCM times a time coordinate 
is not the same as the $\l\to 1$ limit of the plane wave \eqn{ppw}.
In retrospect, this is due to the fact that the zoom in limit is at $\a=0$ whereas the null geodesic leading to \eqn{ppw} passes from $\a=\pi/2$.

\subsection{An alternative plane wave limit of the $\l$-deformed $SU(2)$ }

We add to \eqn{g.su2} the term $-2 k\, dt^2$, we let 
\be
\a=u\, , \qquad t=  \sqrt{1+\l\ov 1-\l}\, u -{1\ov 2\k}  \sqrt{1-\l\ov 1+\l}\, v\,, \qquad \b ={\r\ov \sqrt{2k}}
\ee
and let $k\to \infty$. We obtain a plane wave metric in Rosen coordinates 
\be
\label{gpp.su2}
\begin{split}
&\text{d}s^2=2 du dv + {1-\l^2\ov \D(u)}\sin^2u (dx_1^2 + dx_2^2) \ ,
\\
& \D(u)=(1-\l)^2\cos^2u+(1+\l)^2\sin^2u\,
\end{split}
\ee
and the antisymmetric tensor
\be
\label{Bpp.su2}
B=\Big(\!-u+ {(1-\l)^2\ov  \D(u)}\sin u\cos u\Big)\ dx_1 \wedge dx_2 \, .
\ee
Taking the limit $\l\to 1$, $u\to {1-\l\ov 2} u$  and $x_i\to x_i \sqrt{2/(1-\l)}$
we obtain the solution of the previous subsection.

\no
In Brinkmann coordinates \eqn{gpp.su2} and \eqn{Bpp.su2} read
\be
\begin{split}
& ds^2 = 2 dudv + dx_1^2 + dx_2^2 + F(u)(x_1^2+x_2^2)du^2\ ,
\\
&
F(u)= - {(1-\l)^2\ov \D^2(u)}\Big((1+10\l + \l^2)\cos^2 u+ (1+\l)^2\sin^2u\Big) 
\end{split}
\ee
and 
\be
H= dB= - 2\left( {2\l\ov 1-\l^2} + {1-\l^2\ov \D(u)}\right) du\wedge dx_1\wedge dx_2\ .
\ee
This is a different plane wave background  than that in \eqn{ppw} we mainly consider in this paper.
They become identical only in the CFT limit for $\l=0$. Note that the $\s$-models based on plane wave \eqn{gpp.su2} and \eqn{Bpp.su2} upon fixing the light cone gauge will give rise to masses that depend on the light cone variable $u$ instead of them being constant
as it was the case with the plane wave \eqn{ppw}. 
 
 

\section{The world-sheet metric compatible with $x^+=\tau$}\label{AppB}

\no
The equations of motion of the action \eqn{actga} with respect to the $x^\m$'s are\footnote{For $\g_{ab}=\diag(1,-1)$ and $\e^{01}=1$ we have that $\g_{+-}=\ha$ and that $\e_{+-}=\ha$. Then the equations below
become 
$$
\del_+\del_- x^\m + \Big(\G^\m_{\n\l} - \ha H^\m{}_{\n\l}  \Big) \del_+ x^\n \del_- x^\l = 0\ .
$$
}
\be
\label{kjqq}
\nabla^2 x^\m + \Big(\G^\m_{\n\l} \g^{ab} + \ha H^\m{}_{\n\l} \e^{ab} \Big) \del_a x^\n \del_b x^\l = 0 \ .
\ee
We would like to see if by an appropriate choice of $\g_{ab} $ the equation of motion for $x^-$ admits $x^+=\tau $ as a solution.
To see what such a choice could be, we multiply \eqn{kjqq} with $\sqrt{-\g}$ and we get for $\m= +$ that
\be
\label{delhab}
\del_a h^{a0} + \Big(\G^+_{\n\l}h^{ab} +\ha H^+{}_{\n\l} \bar\e^{ab} \Big)\del_a x^\n\del_b x^\l = 0 \ .
\ee
For the case at hand only $\G^+_{+i}= \ha G^{+-} \del_i G_{+-}$ is non-zero. 
Then 
\be
\begin{split}
& \G^+_{\n\l}h^{ab} \del_a x^\n\del_b x^\l =2  \G^+_{+i} h^{0b}  \del_b x^i= G^{+-} \del_i G_{+-} h^{0b} \del_b x^i  
= G^{+-} \del_b G_{+-} h^{0b}  \ .
\\
\end{split}
\ee
Also we have that
\be
\begin{split}
&\ha H^+{}_{\n\l} \bar\e^{ab} \del_a x^\n \del_b x^\l  =\ha G^{+-}   H_{-ij} \bar\e^{ab} \del_a x^i \del_b x^j
\\
& \phantom{xxxx} =   G^{+-}    (\del_i b_{j-}-\del_j b_{i-}) \del_0 x^i  \del_1 x^j 
\\
&
\phantom{xxxx}  =  G^{+-}  \Big( \del_0(b_{i-}\del_1 x^i)-  \del_1(b_{i-}\del_0 x^i) \Big)\ .
\end{split}
\ee
Therefore  \eqn{delhab}, after multiplying with $G_{+-}$, becomes
\be
\del_a (G_{+-} h^{a0}) +    \del_0(b_{i-}\del_1 x^i) -  \del_1(b_{i-}\del_0 x^i) =0 \ .
\ee
This is solved by 
\be
\label{answ}
h^{00} =  G_{+-}^{-1} \Big(1  - b_{i-} \del_1 x^i \Big) \ ,\qquad h^{01} =   G_{+-}^{-1} b_{i-} \del_0 x^i \ ,
\ee
where we have fixed the integration constants by demanding that for $k\to \infty$ one recovers the conformal gauge $h^{00}=1$ and $h^{01}=0$. Thus, we see that our choice of gauge $x^+=\tau$ is consistent with the equation of motion for $x^-$.

\appendix

\end{document}